\documentstyle[prb,aps,twocolumn,psfig]{revtex}
\begin{document}
\draft
\wideabs{
\title{Low-temperature resistivity of single crystals
YBa$_{2}$Cu$_{3}$O$_{6+x}$ %$
in the normal state }

\author{V.F.Gantmakher\footnote{$^)$e-mail: gantm@issp.ac.ru}$^)$,
L.P.Kozeeva$^{1)}$, A.N.Lavrov$^{1)}$\footnote{$^)$e-mail:
lavrov@casper.che.nsk.su}$^)$, D.A.Pushin, D.V.Shovkun, and G.E.Tsydynzhapov}

\address
{Institute of Solid State Physics RAS, 142432 Chernogolovka, Russia\\
$^{1)}$~Institute of Inorganic Chemistry, Siberian Branch of RAS,
Lavrentyeva-3, 630090 Novosibirsk, Russia}
%$
\maketitle

\begin{abstract}
A scan of the superconductor -- nonsuperconductor transformation in single
crystals YBa$_2$Cu$_3$O$_{6+x}$ ($x\approx0.37$) was done in two alternative
ways, namely, by applying the magnetic field and by reducing the hole
concentration through the oxygen rearrangement. The in-plane normal-state
resistivity $\rho_{ab}$ obtained in both cases was quite similar; its
temperature dependence can be fitted  by logarithmic law in the temperature
range of almost two decades. However, a different representation of the
$\sigma_{ab}=1/\rho_{ab}$ by a power law typical for a 3D-material near a
metal -- insulator transition is also plausible. The vertical conductivity
$\sigma_c=1/\rho_c$ followed the power law and neither $\sigma_c(T)$, nor
$\rho_c(T)$ could be fitted by $\log T$. It follows from the $\rho_c$
measurements that the transformation at $T=0$ is split into two transitions:
superconductor -- normal-metal and normal-metal -- insulator. In our samples,
they are distanced in the oxygen content by $\Delta x\approx0.025$.

\end{abstract} \vspace{0.5cm}

}

In this paper, we intend to apply  the scaling theory of localization \cite{b4}
to the underdoped system YBa$_2$Cu$_3$O$_{6+x}$. The theory \cite{b4} allows
to classify  temperature dependence of the conductivity in the close
vicinity of the metal -- insulator transition in a 3D material. There is
a critical region near the transition where the conductivity $\sigma(T)$
follows a power law

\begin{equation}
\sigma =\alpha +\beta T^{m}.  \label{pow}
\end{equation}
When the main inelastic processes in the critical region are controlled by the
electron-electron interaction, the exponent $m=1/3$\,\, \cite{Im,AA1}. The
constant $\alpha$ in the relation (\ref{pow}) is negative, $\alpha<0$,  and
the critical region there is restricted
from below by a crossover temperature $T^*$ on the
insulating side of the transition. Below $T^*$, the conductivity
falls off exponentially \cite{Mott,AA}

\begin{equation}
\sigma \propto \exp \left[ -(T_{0}/T)^{n}\right] ,\qquad
n=1,\;1/2,\mbox{ or }\;1/4.  \label{exp}
\end{equation}
The crossover temperature $T^*$ and the constant $\alpha$, both become zero
in the transition point so that $\sigma(T)$ in this point is proportional to
$T^{1/3}$\, \,  \cite{Im,AA1}:

\begin{equation}
\sigma =\beta T^{1/3}.  \label{1/3}
\end{equation}
This relation can be used for detecting the transition point.

A low-temperature crossover line exists in the critical region  of the
transition   on the metallic side too.  The conductivity here is described by
dimensionless equation \cite{diagr}

\begin{equation}
s^{3/2}=s^{1/2}+t^{1/2},\quad
s=\sigma(T)/\sigma(0),\quad t=T/T^*.  \label{dimens}
\end{equation}
The function $s(t)$ transforms into power laws (\ref{pow}) in the opposite
temperature limits, with $\alpha =\sigma(0)$ and $m=1/2$ at low temperatures,
when $t\ll1$, and with $\alpha=\frac{2}{3}\sigma(0)$ and $m=1/3$ when
$t\gg1$. It follows from the Eq.(\ref{dimens}) that at the crossover
temperature, when $t=1$, the conductivity is $\sigma(T^*)=1.75\,\sigma(0)$.

According to the theory \cite{b4}, normal-metal -- insulator transition does
not exist in 2D-materials: any film is expected to become insulating at
$T=0$. With decreasing the temperature, the localization starts by the so
called quantum corrections to the classical conductivity $\sigma_0$

\begin{equation}
\sigma =\sigma_0+\Delta\sigma =\sigma_0+\gamma_{\sigma}\log T, \qquad
\Delta\sigma\ll\sigma_0. 				 \label{log}
\end{equation}
Relation $\Delta \sigma<\sigma _{0}$ cannot be violated because the
conductivity $\sigma(T)$ is always positive.  When $\sigma_0$ and
$\Delta\sigma$ become comparable the weak localization turns into the strong
localization and logarithmic behavior (\ref{log}) transforms into
exponential one (\ref{exp}):  at low enough temperature $\sigma(T)$ should
fall off exponentially.

When the metal is superconducting the pattern of the transition into
insulating state changes. In 2D, the transition superconductor -- insulator
has been observed experimentally \cite{Goldman}. In 3D, it is not clear
whether such transition can take place as a single transition or it would
come through intermediate state of normal metal. In order to study
low-temperature behavior of a superconductor, we can bring it  to the
transition, suppress the superconductivity by magnetic field $H$, and then
investigate the transition and its vicinity keeping the field constant. It was
assumed that material with supressed superconductivity would behave at finite
temperatures as an usual metal. However, since 1980,
experimental indications repeatedly appeared that for a superconducting
material there is an intermediate region in the vicinity of the transition in
which the normal resistivity varies logarithmically with the temperature
\cite{Deut}:

\begin{equation}
\rho(T)=\rho_0+\Delta\rho =\rho_0-\gamma_{\rho}\log T. \label{logR}
\end{equation}
The temperature dependence (\ref{logR}) was found in granular aluminum
\cite{Deut} and granular niobium nitride films \cite{cermet}, in percolating
lead films \cite{Gerb}, in Nd$_{2-x}$Ce$_{x}$CuO$_{4-y}$ ceramics
\cite{NdCe}. In all these cases, the resistance changes by several times in
the range of the logarithmic temperature dependence, the logarithmic term in
(\ref{logR}) becoming the leading one at low temperature:

\begin{equation}
\Delta\rho\gg\rho_0.             \label{cond}
\end{equation}
Hence, the relation (\ref{logR}) cannot be converted into (\ref{log}) despite
formal resemblance between them.

The interest to this problem renewed after publications by Ando, Boebinger
et al. \cite{ab1,ab2}. They revealed the $\log T$-term in the resistivity of
the underdoped La$_{2-x}$Sr$_{x}$CuO$_{4}$ in pulsed magnetic fields of 60~T.
This interest has several aspects.\\ (i) As the volume of
experimental data is rather poor one cannot be confident and must check
whether the $\log T$-term really exists ---  it is not simple and even not
always possible to distinguish between $\log T$ and a power-law dependence
(\ref{pow}).\\ (ii) What are the specific properties of those materials where
this term appears. If these were the high-$T_{c}$ superconductors only \cite
{NdCe,ab1,ab2,g1}, it would point to a specific role of strong correlations
in electron systems; on the other hand, it may occur that it is granularity
which is of the primary importance \cite{Deut,cermet,Gerb,NdCe,g1}.

Motivated by these goals, we present below the measurements of the
temperature dependence of the in-plane $\rho_{ab}$ and
vertical $\rho_{c}$ resistance of the single crystals
YBa$_{2}$Cu$_{3}$O$_{6+x}$. By decreasing the doping level in the
high-$T_c$ superconductor systems, one can suppress the superconducting
transition temperature $T_c$ and bring the system to the border of the
superconducting region. In YBaCuO-system, this can be done by decreasing the
oxygen content or, in a limited range, by oxygen rearrangement in the planes
of CuO-chains \cite{Veal,Lavr}.

Single crystals YBa$_2$Cu$_3$O$_{6+x}$ were grown by the flux method in
alumina crucibles \cite{Lavr1}. Being oxygenated at $500^o$C in flowing
oxygen they had $T_c$ of about 90~--~92~K and fairly narrow resistive
transition $\Delta T_c<1$\,K. To bring samples to the border of the
superconducting region, the oxygen content was reduced by high-temperature
($770\div820^o$C) annealing in air with subsequent quenching into liquid
nitrogen \cite{Lavr,Lavr1}. To rearrange the chain-layer oxygen subsystem,
the YBa$_2$Cu$_3$O$_{6+x}$ crystal was heated up to $120\div140^o$C and
quenched into liquid nitrogen. This procedure reduces the mean length of
Cu--O chains; hence, the hole doping of CuO$_2$-planes decreases
\cite{Veal1,Uimin}. The equilibrium state with larger hole density can be
restored simply by room-temperature aging. Dosing the aging time, one can
also obtain intermediate states. So, the quenching--aging procedure allows to
vary gradually the charge carrier density and to tune the sample state
through the border of the superconducting region.

A special care was paid to measure reliably the separate resistivity
components. The in-plane resistivity was measured on thin, 20~--~40~$\mu$m,
plate-like crystals by the four-probe method with current contacts covering
two opposite side-surfaces of the crystal \cite{Lavr1}. The contacts were
painted by silver paste and were fixed by annealing. To measure the vertical
resistivity, the circular current electrodes were painted on the opposite
sides of the plate with potential probes in the middle of the circles. The
resistivity was measured by the standard low-frequency (23 Hz) lock-in
technique in the temperature range 0.37~--~300~K, the measuring current being
low enough to avoid any sign of sample overheating even at the lowest
temperature.

Below we present the temperature dependence of the in-plane resistivity
obtained for the ''aged'' and ''quenched'' states of one of the
YBa$_2$Cu$_3$O$_{6+x}$ ($x\approx0.37$) crystals. For the both states, the
resistivity passed through a minimum near 50~K. The ratio $r$ of the
resistance at room temperature to that at minimum was $\approx\!3$. (In terms
of classical metal physics this means that the crystal is not perfect:
crystals with $r\!\approx\!10$ do exist.) In the quenched state, no signs of
the superconducting transition were observed on the $\rho(T)$ curve down to
the lowest temperature. In the aged state, the resistivity growth at low
temperatures was interrupted by the superconducting transition.  Owing to low
$T_c<10$\,K, the superconducting transition could be suppressed almost
completely by the available magnetic field ${\bf H}\|c$.

\begin{figure}
\vbox{
\psfig{file=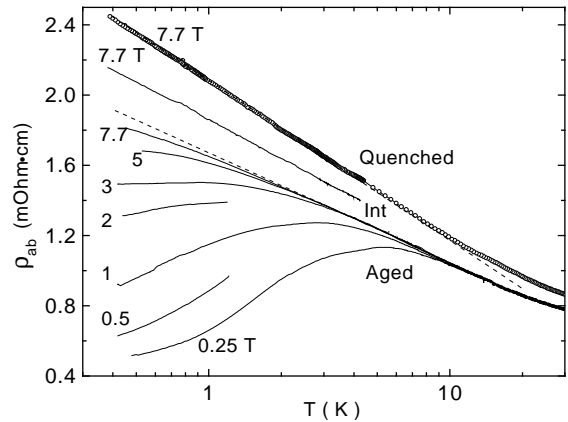,angle=270,width=\columnwidth}

\caption{In-plain resistivity versus $\log T$ for a single crystal YBaCuO with
a fixed content of oxygen but with different its arrangements (quenched,
intermediate and aged states). Only
aged state is superconducting and the set of curves demonstrates how the field
${\bf H}\|c$ destroys the superconductivity. Dashed lines are extrapolation of
the linear dependence $\rho_{ab}(\log T)$. Experimental points are plotted only
on one curve. }  %$
}
\end{figure}

Attempts to fit the $\rho_{ab}(T)$ data by an exponential law (\ref{exp})
were unsuccessful. On the contrary, we succeeded in fitting the data by the
logarithmic law (\ref{log}), see Fig.1. The quenched and the intermediate
states, both without superconductivity, demonstrate the resistivity which
increased logarithmically with decreasing $T$ over almost two decades of
temperature. The magnetoresistance of the quenched state was below 1\,\%; so,
the perfect fit demonstrated in Fig.1 is valid with and without magnetic
field, as well. It can be seen that the $\rho_{ab}(T)$ curves for the aged
state step by step approach with increasing magnetic field the straight line.
Apparently, the deviations from the logarithmic law (\ref{log}) indicate only
that the largest applied field 7.7~T was not strong enough. So, the
representation of the data given in Fig.1 agrees with that of
Refs.\cite{NdCe,ab1,ab2,g1}.

\begin{figure}
\vbox{
\psfig{file=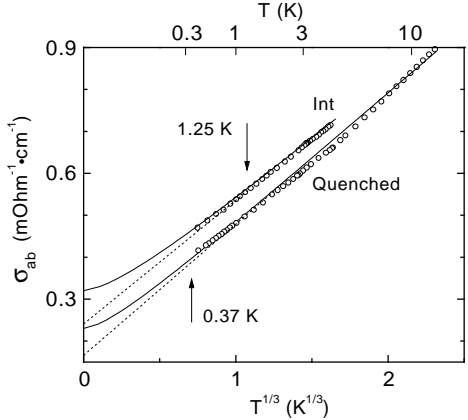,angle=270,width=\columnwidth}
\caption{The data for quenched and intermediate states from Fig.1 replotted
as $\sigma_{ab}$ vs $T^{1/3}$. Solid lines --- fits by Eq.(3) with the
$T^*$ values indicated by arrows, dashed lines --- asymptotes in the $t\gg1$
region. }
%$
}
\end{figure}

However, this is not the only possible interpretation. Assuming that our
sample is a 3D-material, we can analise the data by the help of
Eqs.\,(\ref{pow}) and (\ref{dimens}). According to Fig.2, the data for the
quenched state of the sample replotted as $\sigma$ vs. $T^{1/3}$
approach the straight line at large $T$ and satisfy the equation
(\ref{dimens}) with the values of the parameters
$\sigma(0)=0.23\,$mOhm$^{-1}\cdot$cm$^{-1}$ and $T^*=0.37\,$K.  Those for the
intermediate state are $\sigma(0)=0.32\,$mOhm$^{-1}\cdot$cm$^{-1}$ and
$T^*=1.25\,$K. So, this approach is self-consistent: the aging of the sample
increases the hole density and thus leads to the increase of the conductance
$\sigma(0)$ and of the crossover temperature $T^*$.

So, at this stage we cannot choose between $\log T$-- and $T^m$--
representations, i.e. between representations (\ref{log}) and (\ref{pow}),
(\ref{dimens}). We have made experiments with a crystal with $r\approx10$ too
but were left with the same uncertainty.  However, in any case, the
superconducting state is alternated not by an insulator; representation
(\ref{log}) would indicate that it converts into some specific strongly
correlated metallic state, representation (\ref{pow}) would point to a
normal-metal state. The conclusion that the transformation in YBaCuO consists
of two stages: into normal metal first and into insulator after further
decrease of the hole density, was made previously \cite{NemSh} basing on
extrapolation of the transport data from high temperatures (from above the
resistance minimum). Here the extrapolation edge is far less --- 0.4~K only.

The transport properties of YBaCuO-crystals near the border of the
supercondicting region are highly anisotropic the ratio $\rho_c/\rho_{ab}$
exceeding $10^3$ \cite{anisot}. The crystals can be considered as a stack of
weakly bound conducting CuO$_2$-planes. This brings some uncertainty into the
question whether the in-plane transport should be considered to be of 2D or
of 3D type. At the same time, the vertical transport is certainly 3D and its
temperature dependence is of special interest. We have measured several
crystals and present below examples of typical behavior.

\begin{figure}
\vbox{
\psfig{file=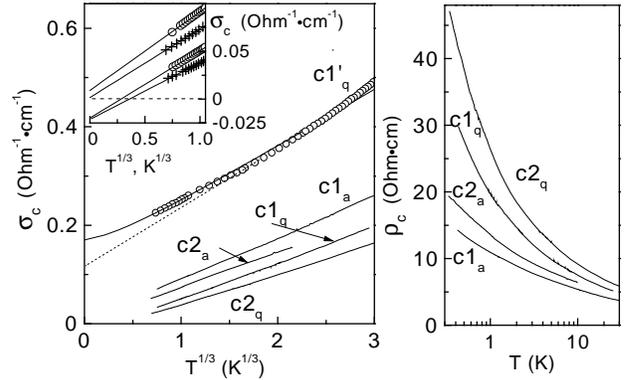,angle=270,width=\columnwidth}
\caption{Vertical resistivity data for two crystals, $c1$ and $c2$, plotted
as $\sigma_c$ vs $T^{1/3}$ (left) and as $\rho_c$ vs $\log T$ (right).
Experimental points are removed from several curves. Insert is enlarged
part of the main plot.}  %$
}
\end{figure}

In Fig.3, the $\rho_c$ data for two crystals annealed in air at $800^o$C,
$c1$ and $c2$, are presented, each in two states, quenched (curves $c1_{\rm
q}$ and $c2_{\rm q}$) and aged ($c1_{\rm a}$ and $c2_{\rm a}$). They
certainly do not follow $\log T$, but perfectly fit the
$T^{1/3}$--representation (\ref{pow}) with $\alpha$ of different signs ---
see inset. Hence, these aged and quenched states should be placed to
different sides of the metal -- insulator transition. All the presented data
were obtained with magnetic field of 7.7~T but the magnetoresistance was
small and did not affect the representation.

In the left part of Fig.3, we present in addition the curve $c1'_{\rm q}$
obtained for the crystal $c1$ annealed at $780^o$C and quenched; the curve is
fitted by eq.(\ref{dimens}) with $\sigma_0=1.5\,\alpha=0.18\,{\rm
Ohm}^{-1}{\rm cm}^{-1}$. The aged state of the crystal with this oxygen
content reveals symptoms of the superconductivity, such as onset kink and
positive magnetoresistance at lower temperatures. Hence, these aged and
quenched states should be placed to different sides of the superconductor --
normal-metal transition on the phase diagram on Figs.\,5 and 7 of
ref.\cite{Lavr}. According to these diagrams, in the $800^o$C range of the
annealing temperature, the $20^o$ change results in $\Delta x\approx0.025$
difference of the oxygen content and in $\Delta n/n_c\approx0.07$ difference
of the hole concentration $n$ normalized to the critical value $n_c$. Precise
positions of both transitions may depend on the degree of disorder in the
crystals. But the obtained numbers can be regarded as estimates for the
distance between the superconductor -- normal-metal and the metal --
insulator transitions along the abscissa axis of the phase diagram.

In conclusion, the low-temperature $\rho_c(T)$ curves of
YBa$_2$Cu$_3$O$_{6+x}$ ($x\approx0.37$) single crystals follow scaling
temperature dependence in the vicinity of the metal -- insulator transition
and allow to specify the transition point. The difference in the oxygen
concentrations $x$ between this point and those of the normal-metal --
superconductor transition is approximately $\Delta x\approx0.025$. It remains
still unclear whether the representation of the in-plane resistivity
$\rho_{ab}(T)$ in the region between these transitions in the $\log T$ scale
is meaningful or the description by the functions (\ref{pow}) and
(\ref{dimens}) is more adequate.

The authors would like to thank A.~Gerber, Y.~Imry, and D.~Khmel'nitskii for
useful discussions. This work was supported by grants RFFI 96-02-17497 and
INTAS-RFBR 95-302 and by the Programs "Superconductivity" and "Statistical
Physics" from the Russian Ministry of Science.

\end{document}